\documentclass{ifacconf}

\usepackage{graphicx}      
\usepackage{natbib}   
\usepackage{comment}
\usepackage{pgfplots}
\pgfplotsset{compat=newest}
\DeclareUnicodeCharacter{2212}{−}
\usepgfplotslibrary{groupplots,dateplot}
\usetikzlibrary{patterns,shapes.arrows}
\usetikzlibrary{shapes.geometric,arrows,positioning}
\pgfplotsset{compat=newest}
\usepackage{float}
\usepackage{algorithm}
\usepackage{algpseudocode}

\usepackage{amsmath}
\usepackage{amssymb}
\newcommand{\bs}[1]{{\ensuremath{\boldsymbol{#1}}}}
\DeclareMathOperator*{\argmin}{\arg\!\min}

\usepackage{cuted}
\usepackage{flushend}

\usepackage{siunitx}
\usepackage{eurosym}

\newcommand{\sectionspace}[1]{
\vspace{-0.2cm}
\section{#1}
\vspace{-0.2cm}}

\newcommand{\subsectionspace}[1]{
\vspace{-0.2cm}
\subsection{#1}
\vspace{-0.2cm}}

\tikzset{font=\small}

\usepackage{xcolor}

\newtheorem{remark}{Remark}

\usepackage[nomain, nonumberlist]{glossaries}
\newglossary{symbols}{sym}{sbl}{List of Symbols}
\glsaddkey{dimension}{\glsentrytext{\glslabel}}{\glsentryd}{\GLsentryd}{\glsd}{\Glsd}{\GLSd}
\glsaddkey{unit}{\glsentrytext{\glslabel}}{\glsentryunit}{\GLsentryunit}{\glsunit}{\Glsunit}{\GLSunit}
\glsaddkey{term}{\glsentrytext{\glslabel}}{\glsentryt}{\GLsentryt}{\glst}{\Glst}{\GLSt}

\newglossaryentry{T}{type=symbols,
	name={\ensuremath{\vartheta_i}},               
	dimension={\ensuremath{1}},
	term={temperature},
	unit = \unexpanded{\si{\degreeCelsius}},
	sort={T},
	description={Mean temperature at day $i$}
}
\newglossaryentry{Tmax}{type=symbols,
	name={\ensuremath{\vartheta_{\text{max},i}}},
	term={maximal temperature},
	dimension={\ensuremath{1}},
	unit = \unexpanded{\si{\degreeCelsius}},
	sort={Tmax},
	description={Maximal temperature at day $i$}
}
\newglossaryentry{Tmin}{type=symbols,
	name={\ensuremath{\vartheta_{\text{min},i}}},
	term={minimal temperature},
	dimension={\ensuremath{1}},
	unit = \unexpanded{\si{\degreeCelsius}},
	sort={Tmax},
	description={Minimal temperature at day $i$}
}
\newglossaryentry{CO2}{type=symbols,
	name={\ensuremath{C_{\text{CO}_\text{2},i}}},
	term={CO\textsubscript{2}-concentration},
	dimension={\ensuremath{1}},
	unit = \unexpanded{\si{ppm}},
	sort={co2},
	description={$\mathrm{CO}_2$ concentration on day $i$}
}
\newglossaryentry{ARID}{type=symbols,
	name={\ensuremath{D_i}},
	term={drought},
	dimension={\ensuremath{1}},
	unit = \unexpanded{\si{1}},
	sort={arid},
	description={Agricultural Reference Index for Drought (ARID) at day $i$}
}
\newglossaryentry{SRAD}{type=symbols,
	name={\ensuremath{R_i}},
	term={radiation},
	dimension={\ensuremath{1}},
	unit = \unexpanded{\si{\mega\joule \per \square\meter \per \day}},
	sort={srad},
	description={Solar radiation (SRAD) at day $i$ \newline
	Typically: $\gls{SRAD}\in\left[0;35\right]\glsunit{SRAD}$}
}

\newglossaryentry{bmass}{type=symbols,
	name={\ensuremath{m_{B,i}}},
	term={biomass},
	dimension={\ensuremath{1}},
	unit = \unexpanded{\si{\kilogram\per\square\meter}},
	sort={bmass},
	description={Biomass at day $i$,  $\left(\SI{10000}{\kilogram\per\hectare}=\SI{1}{\kilogram\per\square\meter}\right)$}
}
\newglossaryentry{Tcum}{type=symbols,
	name={\ensuremath{\tau_i}},
	term={cumulative temperature},
	dimension={\ensuremath{1}},
	unit = \unexpanded{\si{\degreeCelsius\day}},
	sort={Tcum},
	description={Cumulative temperature at day $i$}
}
\newglossaryentry{I50B}{type=symbols,
	name={\ensuremath{I_{50B,i}}},
	term={canopy senescence},
	dimension={\ensuremath{1}},
	unit = \unexpanded{\si{\degreeCelsius\day}},
	sort={I50B},
	description={Cumulative temperature till maturity to reach 50\% radiation interception due to leaf senescence at day $i$}
}
\newglossaryentry{bmass+}{type=symbols,
	name={\ensuremath{m_{\text{B},i+1}}},
	term={biomass},
	dimension={\ensuremath{1}},
	unit = \unexpanded{\si{\kilogram\per\hectare}},
	sort={bmass},
	description={Biomass at day $i$}
}
\newglossaryentry{Tcum+}{type=symbols,
	name={\ensuremath{\tau_{i+1}}},
	term={cumulative temperature},
	dimension={\ensuremath{1}},
	unit = \unexpanded{\si{\degreeCelsius\day}},
	sort={Tcum},
	description={Cumulative Temperature at day $i$}
}
\newglossaryentry{I50B+}{type=symbols,
	name={\ensuremath{I_{50B,i+1}}},
	term={canopy senescence},
	dimension={\ensuremath{1}},
	unit = \unexpanded{\si{\degreeCelsius\day}},
	sort={I50B},
	description={Cumulative temperature till maturity to reach 50\% radiation interception due to leaf senescence at day $i$}
}

\newglossaryentry{fsolar}{type=symbols,
	name={\ensuremath{f_\text{Solar}}},
	term={solar radiation},
	dimension={\ensuremath{1}},
	unit = \unexpanded{1},
	sort={fsolar},
	description={Fraction of solar radiation intercepted by a crop canopy}
}
\newglossaryentry{fheat}{type=symbols,
	name={\ensuremath{f_\text{Heat}}},               
	term={heat stress},
	dimension={\ensuremath{1}},
	unit = \unexpanded{1},
	sort={fheat},
	description={Function for heat stress on biomass growth}
}
\newglossaryentry{fco2}{type=symbols,
	name={\ensuremath{f_{\text{C0}_2}}},               
	term={CO\textsubscript{2} impact},
	dimension={\ensuremath{1}},
	unit = \unexpanded{1},
	sort={fco2},
	description={Function for CO\textsubscript{2} impact on biomass growth}
}
\newglossaryentry{ftemp}{type=symbols,
	name={\ensuremath{f_\text{Temp}}},               
	term={termperature impact},
	dimension={\ensuremath{1}},
	unit = \unexpanded{1},
	sort={ftemp},
	description={Function for temperature impact on biomass growth}
}
\newglossaryentry{fwater}{type=symbols,
	name={\ensuremath{f_\text{Water}}},               
	term={water stress},
	dimension={\ensuremath{1}},
	unit = \unexpanded{1},
	sort={fwater},
	description={Function for water stress on biomass growth}
}

\newglossaryentry{fsolarwater}{type=symbols,
	name={\ensuremath{f_\text{Droght}}},               
	term={radiation impact due to droght stress},
	dimension={\ensuremath{1}},
	unit = \unexpanded{1},
	sort={fsolarwater},
	description={The radiation interception is affected when the drought stress becomes severe enough}
}

\newglossaryentry{HI}{type=symbols,
	name={\ensuremath{I_\text{H}}},
	term={harvest index},
	dimension={\ensuremath{1}},
	unit = \unexpanded{\si{1}},
	sort={HI},
	description={Harvest Index}
}
\newglossaryentry{RUE}{type=symbols,
	name={\ensuremath{E_\text{RU}}},
	term={radiation use efficiency},
	dimension={\ensuremath{1}},
	unit = \unexpanded{\si{\kilogram\per\mega\joule}},
	sort={RUE},
	description={Radiation use efficiency, $\left(\SI{1}{\kilogram\per\mega\joule}=\SI{1000}{\gram\per\mega\joule}\right)$\newline 
	Typically: $\gls{RUE}\in10^{-3}\cdot \left[0.8;2\right]\glsunit{RUE}$ }
}
\newglossaryentry{solarmax}{type=symbols,
	name={\ensuremath{R_\text{max}}},
	term={maximum radiation parameter},
	dimension={\ensuremath{1}},
	unit = \unexpanded{\si{1}},
	sort={fsolarmax},
	description={Maximum fraction of radiation interception that a crop can reach. It is considered as a management parameter, not a crop parameter, to account for different plant spacings. For most high-density crops, this value is set at 0.95.}
}
\newglossaryentry{I50A}{type=symbols,
	name={\ensuremath{I_{50A}}}, 
	term={leaf area parameter},
	dimension={\ensuremath{1}},
	unit = \unexpanded{\si{\degreeCelsius\day}},
	sort={I50A},
	description={Cumulative temperature requirement for leaf area development to intercept 50\% of radiation during canopy closure,}
}
\newglossaryentry{Tsum}{type=symbols,
	name={\ensuremath{\tau_\text{sum}}},
	term={overall cumulative temperature},
	dimension={\ensuremath{1}},
	unit = \unexpanded{\si{\degreeCelsius\day}},
	sort={Tsum},
	description={Cumulative temperature requirement from sowing to maturity }
}
\newglossaryentry{Tbase}{type=symbols,
	name={\ensuremath{\vartheta_\text{base}}},
	term={base temperature},
	dimension={\ensuremath{1}},
	unit = \unexpanded{\si{\degreeCelsius}},
	sort={Tbase},
	description={Base temperature for phenology development and growth}
}
\newglossaryentry{Topt}{type=symbols,
	name={\ensuremath{\vartheta_\text{opt}}},
	term={optimal temperature},
	dimension={\ensuremath{1}},
	unit = \unexpanded{\si{\degreeCelsius}},
	sort={Topt},
	description={Optimal temperature for biomass growth}
}
\newglossaryentry{Theat}{type=symbols,
	name={\ensuremath{\vartheta_\text{heat}}}, 
	term={heat temperature},
	dimension={\ensuremath{1}},
	unit = \unexpanded{\si{\degreeCelsius}},
	sort={Topt},
	description={Threshold temperature to start accelerating senescence from heat stress}
}
\newglossaryentry{Text}{type=symbols,
	name={\ensuremath{\vartheta_\text{ext}}},
	term={extreme temperature},
	dimension={\ensuremath{1}},
	unit = \unexpanded{\si{\degreeCelsius}},
	sort={Topt},
	description={The extreme temperature threshold when \glst{RUE} \gls{RUE} becomes 0 due to heat stress}
}

\newglossaryentry{Sco2}{type=symbols,
	name={\ensuremath{S_{\text{CO}_2}}},
	term={CO\textsubscript{2} sensitivity},
	dimension={\ensuremath{1}},
	unit = \unexpanded{\si{1\per ppm}},
	sort={Sco2},
	description={Relative increase in \glst{RUE} \gls{RUE} per ppm elevated \gls{CO2} above 350 ppm}
}

\newglossaryentry{Swater}{type=symbols,
	name={\ensuremath{S_\text{Water}}},
	term={water sensitivity},
	dimension={\ensuremath{1}},
	unit = \unexpanded{\si{1}},
	sort={Sco2},
	description={Sensitivity of \glst{RUE} \gls{RUE}  (or \glst{HI}) to drought stress (\gls{ARID})}
}

\newglossaryentry{Iwater}{type=symbols,
	name={\ensuremath{I_\text{Water}}},
	term={water stress parameter},
	dimension={\ensuremath{1}},
	unit = \unexpanded{\si{\degreeCelsius}},
	sort={Iwater},
	description={Maximum daily increase in \glst{I50B} \gls{I50B} due to drought stress}
}
\newglossaryentry{Iheat}{type=symbols,
	name={\ensuremath{I_\text{Heat}}},
	term={heat stress parameter},
	dimension={\ensuremath{1}},
	unit = \unexpanded{\si{\degreeCelsius}},
	sort={Iheat},
	description={Maximum daily increase in \glst{I50B} \gls{I50B} due to heat stress}
}
\newglossaryentry{AWC}{type=symbols,
	name={\ensuremath{p_\text{AWC}}},
	term={AWC},
	dimension={\ensuremath{1}},
	unit = \unexpanded{\si{1}},
	sort={AWC},
	description={Fraction of plant available water-holding capacity}
}
\newglossaryentry{RCN}{type=symbols,
	name={\ensuremath{p_\text{RCN}}},
	term={RCN},
	dimension={\ensuremath{1}},
	unit = \unexpanded{\si{1}},
	sort={RCN},
	description={Runoff curve number}
}
\newglossaryentry{DDC}{type=symbols,
name={\ensuremath{p_\text{DDC}}},
term={DDC},
dimension={\ensuremath{1}},
unit = \unexpanded{\si{1}},
sort={DDC},
description={Deep drainage coefficient}
}
\newglossaryentry{RZD}{type=symbols,
	name={\ensuremath{p_\text{RZD}}},
	term={RZD},
	dimension={\ensuremath{1}},
	unit = \unexpanded{\si{\mm}},
	sort={RZD},
	description={Root zone depth }
}

\newglossaryentry{u}{type=symbols,
	name={\ensuremath{\bs{u}_i}},
	term={input vector},
	dimension={\ensuremath{5}},
	sort={u},
	description={\ensuremath{\gls{u}=\begin{bmatrix}\gls{Tmin}\\\gls{Tmax}\\\gls{ARID}\\\gls{SRAD}\\\gls{CO2}\end{bmatrix}}}
}
\newglossaryentry{x}{type=symbols,
	name={\ensuremath{\bs{x}_i}},
	term={state vector},
	dimension={\ensuremath{3}},
	sort={x},
	description={\ensuremath{\gls{x}=\begin{bmatrix}\gls{bmass}\\\gls{Tcum}\\\gls{I50B}\end{bmatrix}}}
}

\newglossaryentry{x_N}{type=symbols,
	name={\ensuremath{\bs{x}_N}},
	term={state vector last day},
	dimension={\ensuremath{3}},
	sort={x_N},
	description={\ensuremath{\gls{x_N}=\begin{bmatrix}\gls{bmass}_N\\\gls{Tcum}_N\\\gls{I50B}_N\end{bmatrix}}}
}
\newglossaryentry{x+}{type=symbols,
	name={\ensuremath{\bs{x}_{i+1}}},               
	term={state vector},
	dimension={\ensuremath{3}},
	sort={x},
	description={\ensuremath{\gls{x}=\begin{bmatrix}\gls{bmass}\\\gls{Tcum}\\\gls{I50B}\end{bmatrix}}}
}
\newglossaryentry{y}{type=symbols,
	name={\ensuremath{\bs{y}_i}},               
	term={output vector},
	dimension={\ensuremath{1}},
	sort={y},
	description={\ensuremath{\gls{y}=\begin{bmatrix}\gls{HI}\cdot \gls{bmass}\end{bmatrix}}}
}
\newglossaryentry{p}{type=symbols,
	name={\ensuremath{\bs{p}}},               
	term={parameter vector},
	dimension={\ensuremath{12}},
	sort={p},
	description={\ensuremath{\gls{p}=\begin{bmatrix}
	\gls{RUE}\\ \gls{solarmax} \\ \gls{I50A} \\ \gls{Tsum} \\ \gls{Tbase} \\ \gls{Topt} \\ \gls{Theat} \\ \gls{Text} \\ \gls{Sco2} \\ \gls{Swater} \\ \gls{Iwater} \\ \gls{Iheat}
\end{bmatrix}}}
}
\newglossaryentry{psoil}{type=symbols,
	name={\ensuremath{\bs{p}_\text{s}}},               
	term={soil parameter vector},
	dimension={\ensuremath{4}},
	sort={psoil},
	description={\ensuremath{\gls{psoil}=\begin{bmatrix}\gls{AWC}\\ \gls{RCN} \\\gls{DDC} \\ \gls{RZD}\end{bmatrix}}}
}
\newglossaryentry{f}{type=symbols,
	name={\ensuremath{\bs{f}}},               
	term={system function},
	dimension={\ensuremath{3}},
	sort={f},
	description={Vector function: $\gls{f}: \mathbb{R}^{\glsd{x}}\times\mathbb{R}^{\glsd{u}}\times\mathbb{R}^{\glsd{p}} \to\mathbb{R}^{\glsd{x}}$}
}

\begin{document}
	\begin{frontmatter}
		
		\title{Optimal Control for Indoor Vertical Farms Based on Crop Growth\thanksref{footnoteinfo}} 

		\author[LSR]{Annalena Daniels\thanksref{equal}} 
		\author[LSR]{Michael Fink\thanksref{equal}} 
		\author[LSR]{Marion Leibold}
		\author[LSR]{Dirk Wollherr}
		\author[DAG]{Senthold Asseng}
		
		\address[LSR]{Chair of Automatic Control Engineering at the Technical University of Munich, Munich, Germany (e-mail: \{a.daniels; michael.fink; marion.leibold; dw\}@tum.de)}
		\address[DAG]{Chair of Digital Agriculture at the Technical University of Munich, Freising, Germany (e-mail: senthold.asseng@tum.de)}
		\thanks[footnoteinfo]{\copyright~2023 the authors. This work has been accepted to IFAC for publication under a Creative Commons Licence CC-BY-NC-ND.}
		\thanks[equal]{Equally contributing authors.}

		\begin{abstract}                
			Vertical farming allows for year-round cultivation of a variety of crops, overcoming environmental limitations and ensuring food security. This closed and highly controlled system allows the plants to grow in optimal conditions, so that it reaches maturity faster and yields more than on a conventional outdoor farm. However, one of the challenges of vertical farming is the high energy consumption. In this work, we optimize wheat growth using an optimal control approach with two objectives: first, we optimize inputs such as water, radiation, and temperature for each day of the growth cycle and second, we optimize the duration of the plant's growth period to achieve the highest possible yield over a whole year. For this, we use a nonlinear, discrete-time hybrid model based on a simple universal crop model that we adapt to make the optimization more efficient. Using our approach, we find an optimal trade-off between used resources, net profit of the yield, and duration of a cropping period, thus increasing the annual yield of crops significantly while keeping input costs as low as possible. This work demonstrates the high potential of control theory in the discipline of vertical farming.
		\end{abstract}
		
		
	\end{frontmatter}
	
	\sectionspace{Introduction}
	
	\vspace{-1.5mm}
	As the world's population continues to grow and is likely to reach more than nine billion people by 2050, the agricultural sector will face ever greater challenges to feed everyone adequately \citep{Searchinger.2019}. Although the amount of food produced must increase, the amount of agricultural land and the energy used must remain the same in order to protect biodiversity and mitigate climate change. At the same time, a large amount of land currently in use will become unusable, for example due to climate change but also due to geopolitical conflicts. Productivity must therefore increase significantly.
	
	\vspace{-0.5mm}
	In order to achieve high yields and quality at minimal cost and environmental impact, research trends show a shift from conventional outdoor agriculture to high-precision controlled environment agriculture (CEA) \citep{Shamshiri.2018}. One step in this direction is the use of greenhouses (GH), which are partially enclosed systems that already control some environmental variables. However, they still typically use external sunlight and temperature, and often do not use fully automated irrigation systems. A step further are the emerging technologies of vertical farms (VF), which come in a variety of forms. However, most have in common that they are completely closed systems and all environmental variables such as light, temperature and water supply are fully controllable. VFs are therefore independent of climatic conditions and at the same time can protect the environment by enabling local production, optimizing processes, and reducing the net area used for cultivation. However, the high initial capital and operating costs of VFs need to be mitigated, as they are still too expensive for most crops. In addition, energy costs must be minimized and yields maximized to increase productivity.
	
	\vspace{-0.5mm}
    VFs are relatively new, but the optimal control of GH climate has been studied extensively. As GHs are partially open systems, not all ambient variables can be optimized in terms of energy and profit. However, since the approach of GH optimization is similar to the one of VFs, it is briefly discussed here. To solve the dynamic economic problem of optimal control in GH environments, most researchers choose a hierarchical control approach based on the work of \citet{vanStraten.2010}. The assumption is that there are two dynamic systems with two different time scales in a GH which allows to divide the optimization into two domains. The lower and faster scale is environment and climate control, for which a model of the GH is needed. The upper and slower scale is plant growth, which is used to calculate the optimal climate setpoints - for GHs, these are mainly the optimal temperature, fertirrigation parameters like the electrical conductivity and/or humidity \citep{vanStraten.2013}. Most of the time only up to 2 parameters and only one objective, the energy cost reduction, are considered. \citet{ramirez2012}, however, also take the water use efficacy and quality of the fruit as an objective into account. The plant growth domain requires an accurate plant model, which will be discussed later. Even if only two parameters are optimized, combining both levels is a complex task \citep{lin.2020}. For this reason, most researchers focus on one of the two, in most cases the faster control of the GH environment, see, e.g., \citet{mahmood.2021}. In approaches where plant growth was also considered (e.g., \citep{rodriguez.2015}), it usually had no direct effect on the optimization because it was not considered in the cost function \citep{su2021greenhouse}. A comprehensive list of control methods for GH climates as well as an overview of energy efficient operation and modeling of GHs can be found in \citep{xu.2019b} and with a special focus on control strategies in \citep{iddio.2020}. However, in order to move to a fully controlled VF, that considers more parameters than temperature and humidity, not only a new climate model is needed, but also a plant growth optimization that considers more aspects than these two and additionally also focuses on a variable growth duration.

    \vspace{-0.2mm}
    Another criticism of existing optimizations is the crop models used. Crop models are used to predict plant growth and to estimate what the particular needs are for the plant at certain stages of development. Existing optimizations use only tomato (\citep{jones1991dynamic} and adaptations) or lettuce crop models \citep{vanHenten1994greenhouse} that are specifically tailored to the problems and are not adapted or tested for other plant species. Whilst many other crop models already exist in the agricultural field, they have been developed in a form that does not easily allow for use in control engineering \citep{vanStraten.2013, engler.2021}. Many single plant models focus on major crops such as wheat \citep{Asseng.2013} and potato \citep{Fleisher.2017}. Models that can be used for multiple crops require a large number of parameters to define each crop. These include the EPIC model \citep{Izaurralde.2006}, the AquaCrop model \citep{Steduto.2009}, and the widely used DSSAT \citep{Jones.2003}. However, often not all of these parameters are available. For this reason, \citet{Zhao.2019} have developed the SIMPLE model, which can be used for a variety of important crops and requires only very few parameters, making it easier to use. For a sustainable future of agriculture, it is crucial to conduct optimization studies for more crop types, which is why the existing models need to be translated or new models need to be developed.

	\vspace{-0.5mm}
	In this paper, we propose and evaluate an open-loop framework for optimizing inputs to a crop growth model in a VF. We consider a VF system as described by \citet{vanDelden.2021}, i.e., an indoor production system without sunlight, where the growing conditions are fully controlled, thus allowing a year-round guarantee of product quantity and quality. The used crop growth model is a state-space representation of the SIMPLE model \citep{Zhao.2019} that we derive. The control objective is to maximize the profit per year, considering the energy costs and harvest yield. As a solution to this optimal control problem (OCP), we compute the ideal input conditions for each day which can be used as setpoints in the VF. In addition, also the optimal duration of crop growth is obtained.
	
    \vspace{-0.5mm}
	Our contributions can be divided into two parts:\\
	\textit{Modelling objectives}: We present a state-space formulation of the SIMPLE crop model in a discrete-time, nonlinear and hybrid form. Furthermore, we derive a smooth and differentiable version of this adapted SIMPLE model.\\
	\textit{Control objectives}: Based on this model, an optimal control algorithm minimizes the inputs of the crop model (drought levels of the soils, temperature, radiation for each day of the growth period) while also maximizing the yield. In a second approach, we add the plant growth period to the optimization variables and obtain the optimal daily inputs for the optimal growth duration. For both approaches, we consider a maturity condition of the plant at harvest.
	
    \vspace{-0.5mm}
	The remainder of this paper is structured as follows:
	In Sec.~2, we introduce the SIMPLE model, propose adaptations, and compare the trajectories of the original to the adapted model. The optimal control framework is introduced in Sec. 3. In Sec. 4, we present results that are discussed in Sec. 5.  Sec. 6 concludes the paper with a summary and an outlook on future work.
	
	\sectionspace{Modeling of crop growth}
	\vspace{-1mm}
	We use the dynamic SIMPLE crop model \citep{Zhao.2019} because it has been carefully calibrated with a large experimental data set for a variety of crops. It can be used for 14 crops and 22 different cultivars by just changing 13 crop parameters, of which four are cultivar parameters and nine are species parameters. As the model is lean and simple, it comes with limitations. For example, it does not consider the effect of vernalization. The effect of soil moisture is included, but nutrient dynamics are not taken into account. 
	In order to apply it to a control problem, we will transform the model into a discrete-time, nonlinear and non-differentiable control system model and adapt it even further to make it differentiable by approximating the discontinuities. 
	The output data of the original model \citep{Zhao.2019} is then used to validate the proposed adaptation.

	\subsectionspace{State-space formulation of the SIMPLE model}
	\vspace{-0.5mm}
	\label{sec:nonsmooth}
	We convert the SIMPLE model \citep{Zhao.2019} into a piece-wise defined nonlinear, discrete-time state-space representation. The state vector for day $i$ is then given as
	\begin{align}
		\gls{x} = \begin{bmatrix} \gls{bmass} & \gls{Tcum} & \gls{I50B}
		\end{bmatrix}^\mathsf{T} ,
	\end{align}
	where $\gls{bmass}$ is the biomass, \gls{Tcum} the cumulative temperature and \gls{I50B} the leaf senescence on day $i$. 

	The inputs for the original SIMPLE crop model \citep{Zhao.2019} are maximum and mean temperature, rainfall, solar radiation, and CO\textsubscript{2} concentration which are typical variables for outdoor cultivation. In contrast to other models, diurnal and nocturnal temperatures are not distinguished and only enter as an average.
    The models that we will derive from this one are meant to be used for indoor farming. Thus, we first make some assumptions about this new system. We assume that the CO\textsubscript{2} concentration in the atmosphere is constant and high, since a preliminary investigation of the system showed a high sensitivity to CO\textsubscript{2}, which meant that the optimum was always at a value of \mbox{$\gls{CO2}=700\ \glsunit{CO2}$}. We also assume a reasonably good temperature control such that there are no temperature peaks during the day, i.e., that the mean temperature and the maximum temperature are the same.
	Under the assumption of a fully controllable environment such as a VF, 
	we choose the simplified system input then as 
	\begin{align}
	\label{eq:inputs}
		\gls{u} = \begin{bmatrix} \gls{T} & \gls{ARID} & \gls{SRAD}
		\end{bmatrix}^\mathsf{T}  ,
	\end{align}
	where \gls{T} is the mean temperature, \gls{ARID} is the relative level of drought \citep{Woli2012}, and \gls{SRAD} is the artificial radiation on day $i$.
	The state-space model (SIMPLE-S) is written as
	\begin{align}\label{eq:sys}
		\gls{x+}&=\gls{x} + \gls{f}(\gls{x},\gls{u}),
	\end{align}
	with \gls{x} the state and \gls{u} the input on day $i$. For a specification of $\gls{f}$, see \citep{Zhao.2019}.
	
	A plant can be harvested when the fruit or the crop is mature. The yield of a mature crop is $ HI \cdot m_{\text{B},N}$, where $HI$ is the harvest index, a parameter of the crop, and $m_{\text{B},N}$ is the biomass on the day of the harvest.
	The maturity of a crop can be determined by 
	\begin{equation}\label{fsolar}
    \begin{aligned}
        \gls{fsolar}(\gls{Tcum}, \gls{I50B}) = \min\left(
		\dfrac{\gls{solarmax}}{1+e^{-0.01\cdot(\gls{Tcum} - \gls{I50A})}}  ,  \right .\\
		\left. \dfrac{\gls{solarmax}}{1+e^{0.01\cdot(\gls{Tcum} - (\gls{Tsum}-\gls{I50B}))}} \right)
	\end{aligned}
	\end{equation} 
	which is a result of combining conditions on cumulative temperature and leaf senescence, as suggested by \citet{Zhao.2019}. The maximum fraction of radiation interception  \gls{solarmax}, the cumulative temperature required for leaf development  \gls{I50A} and the cumulative temperature until maturity \gls{Tsum} are plant specific parameters.

	The crop is mature if 
	\begin{align} \label{eq:terminalConst}
	   \gls{fsolar}(\gls{Tcum}, \gls{I50B}) \leq 0.005
	\end{align}
	and \eqref{fsolar} is decreasing. 
	
	\subsectionspace{Continuous reformulation of the model}\label{ssec:smooth}
	\vspace{-0.5mm}
	As the model definition from \citet{Zhao.2019} is discontinuous and therefore not differentiable, it limits the optimization options drastically as only zero order optimization algorithms without gradients can be used or gradients need to be determined numerically which is computationally expensive and introduces errors. Hence, we also propose a second version of the model: a differentiable form. We adapt the SIMPLE-S \eqref{eq:sys} by smoothing discontinuities to be able to calculate the gradient needed for optimal control.
	The SIMPLE-S model exhibits discontinuities in the form of maximum operators. We use the approximation of the maximum operator proposed by \citet{Biswas2021} and replace all maximum operators in~\gls{f}. The approximation of the maximum operator is
	\begin{align}
	\label{eq:maxop}
		\max(a,b)\approx\operatorname{smax}_\epsilon(a,b)=\frac{a + b + \sqrt{(a-b)^2 + \epsilon } }{2},
	\end{align}
	where $\epsilon$ is a positive constant.  

	The continuous crop model will be called SIMPLE-SC and written as
	\begin{align}\label{eq:sysS}
		\gls{x+}&=\gls{x} + \gls{f}_\text{s}(\gls{x},\gls{u}).
	\end{align}
	For small $\epsilon$, the smooth maximum function converges to the standard maximum operator. Thus for $\epsilon$ = 0, ${\gls{f}_\text{s}=\gls{f}}$ and \eqref{eq:sysS} becomes \eqref{eq:sys}. 
	\subsectionspace{Model validation}
	\vspace{-0.5mm}
	In order to use the SIMPLE-S \eqref{eq:sys} and SIMPLE-SC \eqref{eq:sysS} for optimization, we first compare their behavior with the original results of the SIMPLE model \citep{Zhao.2019} as reference. 
	The comparison was made for all available crops and cultivars. 
	Resulting trajectories for wheat of the 'Batten' cultivar are shown in  Fig.~\ref{fig:modelver} with different values for~$\epsilon$. We note that a sufficiently small value for $\epsilon$ must be chosen in order to achieve a good approximation of the reference model. Small values for $\epsilon$ mean that the approximation equations of the individual non-smooth functions are very close to the original course of the equations. This can have the disadvantage that the derivatives or gradients can become very large or small, which can have an unfavorable effect on the optimization. For the following optimization, we choose $\epsilon = 10^{-4}$ as a trade-off. 
	
	\newsavebox{\modelver}
	\begin{lrbox}{\modelver} 
	\input{figs/modelverification}
	\end{lrbox}
	\begin{figure}[ht]
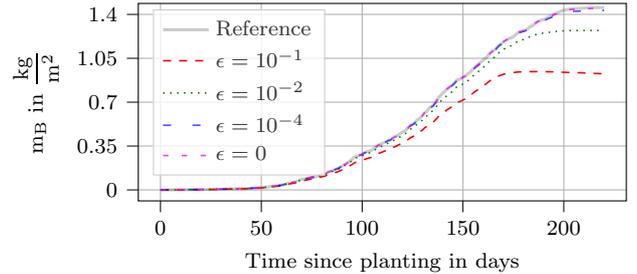

		\centering 
		\usebox{\modelver}
		\caption{Trajectories of the biomass of the SIMPLE-SC \eqref{eq:sysS} with different parameters~$\epsilon$ and a reference trajectory given by the SIMPLE model \citep{Zhao.2019}. For sufficiently small $\epsilon$, the model approaches the reference.} 
		\label{fig:modelver}
	\end{figure}  
	
 	\sectionspace{Optimal Control}
 	\vspace{-1mm}
	In the following, the framework for the optimal control of plant growth is presented.  First, the growth period is specified, i.e., the final time of the OCP is fixed. The inputs to the system are computed based on a cost function that enforces small inputs and high yields. However, in a VF, multiple harvests per year are possible. Therefore, in the second part, the additional question arises how long the crop should grow to achieve the maximum yield per year, i.e., the growing period is a free final time of the OCP.
	
	\subsectionspace{Optimal control with fixed final time}
	\vspace{-0.5mm}
	The optimization is based on a cost function 
	\begin{align}\label{eq:cost}
		J(\bs{U},\bs{x}_0) = \sum_{i=0}^{N-1}  l(\gls{u}) - V(\bs{x}_N)
	\end{align}
	that is defined over a given growth period of $N$ days. The input sequence is denoted as $\bs{U} = \left[\bs{u}_0^\mathsf{T}, \bs{u}_1^\mathsf{T}, ..., \bs{u}_{N-1}^\mathsf{T}\right]^\mathsf{T}$.
	The stage cost $l(\gls{u})$ represents the energy cost used for the growing process and is given as $l(\gls{u}) =\gls{u}^\mathsf{T}  \bs{R} \gls{u} + \bs{r}^\mathsf{T} \gls{u}$, where $\bs{R} \in \mathbb{R}^{3\times 3}$ and $\bs{r} \in \mathbb{R}^{3}$ are weights that will be specified later. The terminal cost $V(\bs{x}_N)$ gives the yield at the harvest on day $N$. The price evolves linearly, leading to a linear cost term, i.e.  $V(\bs{x}_N) =\bs{q}^\mathsf{T} \gls{x_N}$, with the weight $\bs{q}\in\mathbb{R}^{3}$. The cost function \eqref{eq:cost} represents the negative economic yield, thus the terminal cost is used with a negative sign. 
	Temperature, water, and radiation can be controlled in a VF, but their values are bounded. Therefore, we add constraints $\gls{u}\in \mathcal{U}$ to the OCP.
	
	These considerations yield the OCP
	\begin{subequations}\label{eq:opt}
		\begin{align}
			\bs{U}^* =& \argmin_{\bs{U}}  J(\bs{U},\bs{x}_0)  \\
			\text{s.t. }& \gls{x+}=\gls{x} + \gls{f}_\text{s}(\gls{x},\gls{u})  &\forall i \in[0,N-1] \\
			& \gls{u} \in \mathcal{U}  \quad &\forall i \in[0,N-1] \\
			& g(\bs{x}_N) = 0 \\
			& \bs{x}_0 = \bs{x}_\text{init},
		\end{align}
	\end{subequations}
	where $\bs{U}^*$ is the sequence of optimal inputs. The states on day $i = 0$ are given as initial states $\bs{x}_\text{init}$.  
	The system model $\gls{f}_\text{s}$ refers to the differentiable SIMPLE-SC \eqref{eq:sysS}. 
	The function $g(\bs{x}_N)$ guarantees maturity of the plant before harvest, as it enforces the condition \eqref{eq:terminalConst} with equality, i.e., $g(\bs{x}_N)= \gls{fsolar}(\gls{Tcum}, \gls{I50B}) - 0.005$.
	
	\subsectionspace{Optimal control with free final time} \label{ssec:timeopt}
	\vspace{-0.5mm}
    In this section, the optimization \eqref{eq:opt} is extended to an OCP with free final time. After the harvest, it is assumed that new seeds can be planted the next day and that a permanent cultivation in the VF can be achieved throughout the year, which allows to optimize the economic outcome over this time. The input parameters can be chosen in a way that the crop either takes longer to ripen or can reach maturity quickly. As can be seen in Fig. \ref{fig:Cropinayear}, a quick growth of the crop leads to a higher annual yield, but will at the same time also lead to higher energy costs. Introducing the growth period as an additional decision variable creates a new OCP. 
	\newsavebox{\cropinayear}
	\begin{lrbox}{\cropinayear} 
	\input{figs/CropinaYear}
	\end{lrbox}
	\begin{figure}[ht]
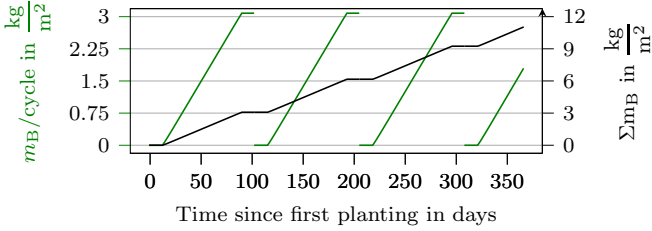

		\centering
		\usebox{\cropinayear}
		\caption{Wheat growth over a year in a permanent cultivation. The highest possible biomass per growth cycle (left scale, green) is limited by the crop itself but the number of harvests in a VF can be more than 1. It directly co-relates with the length of the growth period. Thus, the cumulative biomass and therefore yield in a year (right scale, black) can be much higher if multiple cycles are achieved in a year.}
		\label{fig:Cropinayear}
	\end{figure}
	
	If the number of days is free in the OCP, then the length of the input sequence vector \bs{U} is affected, which leads to discontinuities in the optimization. However, one can note that the \mbox{SIMPLE-SC} \eqref{eq:sysS} has the same structure as an Euler discretization. Thus, to solve the problem of the changing length of the input vectors, we do not work with a variable $N$, but introduce a variable sampling time $T$ into the system dynamics, resulting in an updated version of the system (SIMPLE-SCS) given as
	\begin{align}\label{eq:sysT}
		\gls{x+}&=\gls{x} + T\gls{f}_\text{s}(\gls{x},\gls{u}).
	\end{align}
	The sampling time in the original system is set to $T=1$. That means, one evaluation of the system \eqref{eq:sysS} computes the state of the next day. With \eqref{eq:sysT}, other sampling times such as half a day ($T=0.5$) can be used to change the growing duration.  
	However, the Euler discretization is an approximation of a derivative. It follows that a system with a sampling time $T\neq 1$ is an approximation and is subject to deviations from model \eqref{eq:sysS} and accurate results can only be obtained if $T=1$. 

	The cost function \eqref{eq:cost}, which is defined for one growing cycle, is adjusted to consider one year, i.e.,
	\begin{align}\label{eq:costT}
		J_\text{t}(\bs{U}_T,\bs{x}_0) = \sum_{i=0}^{N-1} \frac{365}{N}l(\gls{u}) - \frac{365}{N T}V(\bs{x}_N),
	\end{align}
	where $\bs{U}_T$ is given as the optimal input vector $\bs{U}$ extended by $T$, i.e. $\bs{U}_T = \left[\bs{U}, T\right]$.
	The factor $\frac{365}{N}$ represents how often the plant can be harvested within one year. As the sampling time $T$ decreases, more harvests are possible in a year, allowing for a higher yield. Therefore, the sampling time $T$ also affects the terminal cost $V(\bs{x}_N)$. The stage cost $l(\gls{u})$ is not affected because the factor $T$ would show up in both the daily cost and the harvest frequency, thus truncating it.
	The model \eqref{eq:sysT} and the cost function \eqref{eq:costT} are combined in the OCP
	\begin{subequations}\label{eq:optT}
		\begin{align}
			\bs{U}_T^* &= \argmin_{\bs{U}_T}  J_\text{t}(\bs{U}_T,\bs{x}_0)  \\
			\text{s.t. }& \gls{x+}=\gls{x} +T \gls{f}_\text{s}(\gls{x},\gls{u})  &\forall i \in[0,N-1] \\
			& \gls{u} \in \mathcal{U}  \quad &\forall i \in[0,N-1] \\
			& T \in [0.5;1.5]\\
			& g(\bs{x}_N) = 0 \\
			& \bs{x}_0 = \bs{x}_\text{init},
		\end{align}
	\end{subequations}
	in which the sampling time is also a decision variable. 
	After determining an optimized sampling time $T^*$, the optimal duration of one growing period is 
	\begin{align}
		N^* = \lfloor  T\cdot N \rfloor .
	\end{align}
	Only integer values are significant for the length of the growth period, hence $N^*$ is finally rounded down. 
	\begin{remark}
	Due to the approximation of the growth dynamics, an accurate result cannot be obtained if the optimization yields an optimal $T \neq 1$ and $N^* \neq N_0$, where $N_0$ is the initial value for the length of the growth period. Thus, the optimization must be repeated with the adjusted growth period $N$ until the optimization yields a $T$ sufficiently close to 1.
    \end{remark}
    
    \vspace{-0.5mm}
	We propose an iterative approach to optimization in this paper, adjusting $N$ in each iteration. The process ends when $T$ is close to 1 and thus the model is again accurate. For this purpose, the threshold~$\delta$ is used, i.e., $|T-1|<\delta$. The method is summarized in Algorithm~\ref{alg}.
	
	\begin{algorithm}[H]
		\caption{Optimization with free final time}\label{alg}
		\begin{algorithmic}
			\State $N \gets N_0$
			\While{$|T-1|>\delta$}
			\State Solve \eqref{eq:optT}
			\State $T \gets T^*$
			\State $N \gets \lfloor  T\cdot N \rfloor $
			\EndWhile
		\end{algorithmic}
	\end{algorithm}
	
	\vspace{-0.5mm}
	\subsectionspace{Objective and weights of the cost functions}
	\vspace{-0.5mm}
	A reasonable trade-off between crop economic yield and energy consumption needs to be obtained. Economic profit is represented with a negative target cost in \eqref{eq:cost} and \eqref{eq:costT}. They are calculated from the difference between the energy costs and the income value of the yield.
	The costs of heating, radiation, and irrigation are considered in the cost function by choosing appropriate weights, which we propose as 
	\begin{align}\label{eq:inputweigt}
		\ensuremath{\bs{R}} = \begin{bmatrix}
			C_{\vartheta} &0 &0\\
			0 &C_{D} &0 \\
			0 & 0 &0
		\end{bmatrix},
		\quad 
		\ensuremath{\bs{r}} = \begin{bmatrix}
			-2 C_{\vartheta}\vartheta_0 &-2 C_{D} &C_R
		\end{bmatrix}.
	\end{align} 
	for the cost functions in \eqref{eq:cost} and \eqref{eq:costT}.
	Light is produced by LEDs and the constant $C_R$ is the price to generate \SI{1}{\mega\joule} of plant available radiation. The ambient temperature $\vartheta_0$ would not produce costs, however, heating and cooling consumes electricity. Thus, we use a quadratic function $C_{\vartheta}(\gls{T}-\vartheta_0)^2$ for the cost to change the temperature, where $C_{\vartheta}$ is a constant. 
	The drought index \gls{ARID} is 1 for dry soil which corresponds to zero cost. If the drought index \gls{ARID} is zero, water is available which affects the cost. 
	We propose the quadratic term $C_D(\gls{ARID}-1)^2$ to model this cost where $C_D$ is a constant. 

	The terminal cost is evaluated on the day of the harvest. For the computation of the yield, only the biomass and the crop related harvest index $HI$ are~relevant 
	\begin{align}
		\ensuremath{\bs{q}}^\mathsf{T} = \begin{bmatrix}
			HI  c_\text{crop} &0 &0
		\end{bmatrix},
	\end{align}
	where $c_\text{crop}$ is the price of one weight unit of the crop.  
	
	\sectionspace{Results}
	\vspace{-1mm}
	An algorithm that solves the OCPs~\eqref{eq:opt} and \eqref{eq:optT} is implemented in Python. It includes an automatic differentiation approach provided by \citet{Andersson2019}. We show the results for the wheat cultivar "Batten" as an example of the crop model as wheat is a crop that has not attracted much attention in VFs, but may have high potential if the feasibility gap of the cost-yield ratio can be bridged \citep{asseng2020}.
	Inputs are bounded for all days~$i$ to $\gls{T} \in [0;35]\glsunit{T}$, ${\gls{ARID}\in[0;1]}$, and $\gls{SRAD} \in[0;35]\glsunit{SRAD}$.
	The \Glst{CO2} is not optimized and is set to \mbox{$\gls{CO2}=700\ \glsunit{CO2}$}.
	We choose the parameters in \eqref{eq:inputweigt} as $C_{\vartheta} =\ 1.8 \cdot 10^{-6}  \frac{\geneuro}{\glsunit{T}^2}$, $C_D = 0.02\ \geneuro$, ${C_R = 0.038\ \frac{\geneuro}{(\glsunit{SRAD})}}$, and $\vartheta_0 = 10\ \glsunit{T}$, based on current average conditions in the world.
	The harvest index for "Batten" is $HI=0.3$ and we choose $c_\text{crop} = 132.9\ \frac{\geneuro}{kg}$ as the notional wheat price, which is 400 times higher than current market prices. This high price is needed to simulate profitability, which is not yet given for wheat in VFs \citep{asseng2020}.
	The results of our approach are presented in three steps: First, we show the results of the OCP with a fixed final time, then with a free final time, and finally we compare both approaches with each other and with scenarios without any optimization.

	\subsectionspace{Optimal control with fixed final time}
	\vspace{-0.5mm}
	\label{sec:fixedfinaltime}
	First, we only consider an input optimization for the variables \gls{SRAD}, \gls{ARID}, and \gls{T} for all $i\in[0;N-1]$. In this case, a reasonable length of the growth period $N$ needs to be guessed and is set to $N=102$ days in this example.
	By solving the OCP \eqref{eq:opt}, we find an input trajectory (see Fig.~\ref{fig:inputs}) which maximizes the yield of the biomass and minimizes the energy costs. It is noticeable that all input trajectories remain constant most of the time. In the middle of the growth period, it is cost-optimal to always provide the plant with sufficient water ($D = 0$), to keep the temperature \gls{T} at the optimal temperature of $\gls{Topt} = 15\ \glsunit{T}$ for wheat, and to set the irradiation \gls{SRAD} to the constrained maximum of 35\ \glsunit{SRAD}. Only in the beginning does a higher temperature \gls{T} lead to such an acceleration of germination and growth that it compensates for the higher costs. Towards the end, when the plant needs to mature and dry, less water and less radiation, but again a higher temperature is needed. The obtained optimal trajectory of the states that can be seen in Fig. \ref{fig:states}.
	\newsavebox{\inputs}
	\begin{lrbox}{\inputs} 
\begin{tikzpicture}

\definecolor{darkgray176}{RGB}{176,176,176}
\definecolor{lightgray204}{RGB}{204,204,204}
\definecolor{orange}{RGB}{255,165,0}

\begin{axis}[
height=4cm,
legend cell align={left},
legend style={
  fill opacity=0.8,
  draw opacity=1,
  text opacity=1,
  at={(0.03,0.97)},
  anchor=north west,
  draw=lightgray204
},
minor xtick={},
minor ytick={},
tick align=outside,
tick pos=left,
width=7cm,
x grid style={darkgray176},
xlabel={Time since planting in days},
xmajorgrids,
xmin=-5.05, xmax=106.05,
xtick style={color=black},
xtick={-20,0,20,40,60,80,100,120},
y grid style={darkgray176},
ylabel={\(\displaystyle \vartheta \mathrm{\ in \ } ^o C,  R \mathrm{\ in \ } \frac{\mathrm{MJ}}{\mathrm{m}^2\mathrm{d}}\)},
ymajorgrids,
ymin=-1.75, ymax=36.75,
ytick style={color=black},
ytick={0,7,14,21,28,35}
]
\addplot [semithick, red]
table {%
0 34.9999974507166
1 34.9999974507168
2 34.999997450717
3 34.9999974507174
4 34.9999974507179
5 34.9999974507186
6 34.9999974507195
7 34.9999974507207
8 34.9999974507223
9 34.9999974507243
10 34.9999974507264
11 34.999997450728
12 34.9999974507259
13 33.8549984505361
14 33.8050714034942
15 33.7281781222744
16 33.5665889283532
17 25.2314406126902
18 16.0730363434641
19 15.7747186615667
20 15.6511129933636
21 15.5815487443869
22 15.5368012474887
23 15.5058049607489
24 15.4833260279025
25 15.466521984876
26 15.4536935420741
27 15.4437501063997
28 15.4359531229596
29 15.429781873113
30 15.4248582692249
31 15.420902003806
32 15.4177021932038
33 15.4150984862367
34 15.4129679346748
35 15.4112155871377
36 15.40976761695
37 15.4085662410889
38 15.4075659357717
39 15.4067306034532
40 15.4060314429561
41 15.405445341414
42 15.4049536547025
43 15.4045412780709
44 15.4041959344511
45 15.4039076268949
46 15.4036682155662
47 15.4034710900103
48 15.4033109150177
49 15.4031834340199
50 15.4030853181418
51 15.4030140521751
52 15.4029678511346
53 15.4029456029203
54 15.4029468341076
55 15.402971697141
56 15.4030209783298
57 15.4030961271201
58 15.4031993082301
59 15.4033334794886
60 15.4035024996945
61 15.4037112726714
62 15.4039659360683
63 15.4042741066059
64 15.4046451976642
65 15.4050908307878
66 15.4056253704046
67 15.4062666215765
68 15.4070367449854
69 15.4079634629827
70 15.4090816572896
71 15.4104354953733
72 15.4120812722625
73 15.4140912232929
74 15.4165586611568
75 15.4196049383228
76 15.4233889781202
77 15.4281205520875
78 15.4340793145135
79 15.4416432629401
80 15.4513336473169
81 15.4638902814874
82 15.4804062329988
83 15.5025862608145
84 15.5332870012982
85 15.5777840560345
86 15.6472856266166
87 15.7716511989975
88 16.0767691653064
89 28.7027032491304
90 29.0187477547493
91 29.0187491053629
92 29.018751050239
93 29.018753380295
94 29.0187556397791
95 29.0187574226627
96 29.0187586099683
97 29.0187593118651
98 29.0187596973812
99 29.0187599005042
100 29.0187600051702
101 29.0187600584821
};
\addlegendentry{$\vartheta$}
\addplot [semithick, orange]
table {%
0 7.37744078004821e-08
1 7.59872482244104e-08
2 7.91405431291849e-08
3 8.3645361155016e-08
4 9.01042751723387e-08
5 9.94130757934805e-08
6 1.12929967162884e-07
7 1.32771924481858e-07
8 1.62368634791365e-07
9 2.07586315433653e-07
10 2.7926092221339e-07
11 3.9977432557978e-07
12 6.23977925857636e-07
13 34.9999955220437
14 34.999999087974
15 34.9999995453802
16 34.9999997180293
17 34.999999804853
18 34.9999998430492
19 34.9999998610177
20 34.9999998752033
21 34.9999998866111
22 34.9999998958646
23 34.9999999034078
24 34.9999999095731
25 34.9999999146183
26 34.9999999187478
27 34.9999999221274
28 34.9999999248928
29 34.9999999271561
30 34.9999999290101
31 34.999999930531
32 34.9999999317812
33 34.9999999328116
34 34.9999999336632
35 34.9999999343692
36 34.9999999349563
37 34.9999999354459
38 34.9999999358552
39 34.9999999361982
40 34.9999999364861
41 34.9999999367281
42 34.9999999369316
43 34.9999999371027
44 34.9999999372464
45 34.9999999373667
46 34.999999937467
47 34.99999993755
48 34.9999999376179
49 34.9999999376726
50 34.9999999377153
51 34.9999999377473
52 34.9999999377693
53 34.9999999377818
54 34.9999999377851
55 34.9999999377794
56 34.9999999377642
57 34.9999999377393
58 34.999999937704
59 34.9999999376571
60 34.9999999375974
61 34.9999999375232
62 34.9999999374324
63 34.9999999373223
64 34.9999999371898
65 34.9999999370308
66 34.9999999368404
67 34.9999999366127
68 34.9999999363404
69 34.9999999360143
70 34.9999999356232
71 34.9999999351531
72 34.9999999345866
73 34.999999933902
74 34.9999999330719
75 34.9999999320624
76 34.9999999308307
77 34.9999999293237
78 34.9999999274757
79 34.9999999252056
80 34.9999999224146
81 34.9999999189822
82 34.9999999147618
83 34.999999909574
84 34.999999903196
85 34.9999998953468
86 34.9999998856629
87 34.9999998736621
88 34.9999998586901
89 34.9999998400376
90 2.42988930477429e-07
91 2.20192859882175e-07
92 1.88829929040069e-07
93 1.53766194091775e-07
94 1.22496367671007e-07
95 9.97363904770831e-08
96 8.55010521252517e-08
97 7.74241824749558e-08
98 7.30929907474071e-08
99 7.08405298141806e-08
100 6.96877875061136e-08
101 6.91026960740887e-08
};
\addlegendentry{$R$}
\end{axis}

\begin{axis}[
axis y line=right,
height=4cm,
legend cell align={left},
legend style={
  fill opacity=0.8,
  draw opacity=1,
  text opacity=1,
  at={(0.03,0.03)},
  anchor=south west,
  draw=lightgray204
},
minor xtick={},
minor ytick={},
tick align=outside,
width=7cm,
x grid style={darkgray176},
xmin=-5.05, xmax=106.05,
xtick pos=left,
xtick style={color=black},
xtick={-20,0,20,40,60,80,100,120},
y grid style={darkgray176},
ylabel={\(\displaystyle D \mathrm{\ in\ } \%\)},
ymin=-5, ymax=105,
ytick pos=right,
ytick style={color=black},
ytick={0,20,40,60,80,100},
yticklabel style={anchor=west}
]
\addplot [semithick, blue]
table {%
0 5.41771320016197
1 5.4177132001629
2 5.41771320016422
3 5.41771320016603
4 5.4177132001686
5 5.41771320017202
6 5.41771320017667
7 5.4177132001827
8 5.41771320019037
9 5.41771320019942
10 5.41771320020843
11 5.41771320021115
12 5.41771320018653
13 4.45303883388427e-07
14 3.62807508292073e-07
15 3.00838755241827e-07
16 2.51673204748378e-07
17 1.99114750032179e-07
18 1.88229302805285e-07
19 1.8301718939684e-07
20 1.78664751113233e-07
21 1.74997854541852e-07
22 1.71906522373876e-07
23 1.6930482103197e-07
24 1.67121338632348e-07
25 1.6529505046997e-07
26 1.63772976257944e-07
27 1.62508735787178e-07
28 1.61461665220663e-07
29 1.60596282918339e-07
30 1.59881929696764e-07
31 1.59292447615648e-07
32 1.58805818383401e-07
33 1.5840374016782e-07
34 1.58071161292874e-07
35 1.57795805249092e-07
36 1.57567719456129e-07
37 1.57378869797929e-07
38 1.57222791461857e-07
39 1.57094297700612e-07
40 1.56989242628099e-07
41 1.56904331508137e-07
42 1.5683697126679e-07
43 1.56785154329098e-07
44 1.56747369770222e-07
45 1.56722536826823e-07
46 1.56709956850887e-07
47 1.56709280723909e-07
48 1.56720489560239e-07
49 1.5674388722316e-07
50 1.56780103779569e-07
51 1.56830109550392e-07
52 1.56895239913058e-07
53 1.56977231497515e-07
54 1.57078270930729e-07
55 1.5720105785017e-07
56 1.57348884571248e-07
57 1.5752573559707e-07
58 1.57736411166706e-07
59 1.57986680319775e-07
60 1.58283470616443e-07
61 1.58635103819392e-07
62 1.59051589693268e-07
63 1.59544993836937e-07
64 1.6012990042348e-07
65 1.60823997261206e-07
66 1.6164881915724e-07
67 1.62630696688008e-07
68 1.63801971691765e-07
69 1.65202558517315e-07
70 1.66881951325255e-07
71 1.6890180177167e-07
72 1.71339215982508e-07
73 1.74290940524018e-07
74 1.77878617648052e-07
75 1.82255284068197e-07
76 1.87613265828058e-07
77 1.94193608016152e-07
78 2.02297240924054e-07
79 2.12298354746232e-07
80 2.24661110354294e-07
81 2.39962010503791e-07
82 2.58922061625842e-07
83 2.82455326579898e-07
84 3.11744003385505e-07
85 3.48356133970038e-07
86 3.94433761626229e-07
87 4.53004559992827e-07
88 5.28531382318204e-07
89 6.31437284733441e-07
90 99.9998930859288
91 99.9998930757431
92 99.9998930610825
93 99.9998930437039
94 99.9998930271809
95 99.9998930144331
96 99.9998930060996
97 99.9998930012334
98 99.9998929985801
99 99.9998929971879
100 99.9998929964723
101 99.9998929961083
};
\addlegendentry{D}
\end{axis}

\end{tikzpicture}
	\end{lrbox}
	\begin{figure}[ht]
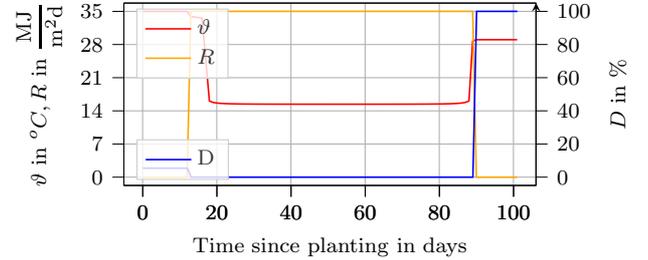

		\centering
		\usebox{\inputs}
		\caption{Optimal input trajectories of temperature \gls{T}, radiation \gls{SRAD} (both left scale), and drought \gls{ARID} (right scale) for an input optimization with a fixed final time of $N = 102$.}
		\label{fig:inputs}
	\end{figure}
	
    The development of biomass \gls{bmass}, i.e. the growth of the plant, depends not only on the inputs, but also on $\gls{fsolar}$ \eqref{fsolar} and thus on the other two states \gls{Tcum} and \gls{I50B}. The optimal growth is achieved by keeping the value of $\gls{fsolar}(\gls{Tcum}, \gls{I50B})$ close to its maximum value for a long time (see Fig. \ref{fig:fsolar}). Since \gls{fsolar} depends only on the cumulative temperature \gls{Tcum} in the growth period, \gls{Tcum} is first steered to the value at which \gls{fsolar} is maximum (see Fig.~\ref{fig:states}). The value of $\gls{fsolar}$ remains constant then by keeping the temperature low (see Fig. \ref{fig:inputs}). However, lower temperatures reduce biomass growth. 
	At the end of the growth period, \gls{fsolar} also depends on \gls{I50B} (see \eqref{fsolar}) and the terminal condition must be satisfied, which means that the sum of \gls{I50B} and \gls{Tcum} must become large. This is achieved by increasing the temperature again, which affects both conditions. Drought \gls{ARID} also goes up to increase \gls{I50B}. Radiation \gls{SRAD} is decreased to reduce the stage cost $l(\gls{u})$. 
	
	\newsavebox{\states}
	\begin{lrbox}{\states}
\begin{tikzpicture}

\definecolor{darkgray176}{RGB}{176,176,176}
\definecolor{green}{RGB}{0,128,0}
\definecolor{lightgray204}{RGB}{204,204,204}
\definecolor{magenta}{RGB}{255,0,255}

\begin{axis}[
height=3.5cm,
legend cell align={left},
legend style={
  fill opacity=0.8,
  draw opacity=1,
  text opacity=1,
  at={(0.03,0.97)},
  anchor=north west,
  draw=lightgray204
},
minor xtick={},
minor ytick={},
tick align=outside,
tick pos=left,
width=7cm,
x grid style={darkgray176},
xmajorgrids,
xmin=-5.1, xmax=107.1,
xtick style={color=black},
xtick={-20,0,20,40,60,80,100,120},
y grid style={darkgray176},
ylabel={\(\displaystyle {\mathrm{m_\mathrm{B}\ in \ }}{\mathrm{\frac{kg}{m^2}}}\)},
ymajorgrids,
ymin=-0.1875, ymax=3.1875,
ytick style={color=black},
ytick={0,0.75,1.5,2.25,3}
]
\addplot [semithick, green]
table {%
0 0
1 4.50002165436748e-12
2 1.09131716944179e-11
3 2.00726011073812e-11
4 3.3191995713209e-11
5 5.20536602525357e-11
6 7.93002869706836e-11
7 1.1889690035792e-10
8 1.76885099161156e-10
9 2.62675046077529e-10
10 3.9142160205327e-10
11 5.88847022572664e-10
12 9.02432890122521e-10
13 1.43290747952325e-09
14 0.0345129874377385
15 0.0705863648402166
16 0.107860719276478
17 0.146040322795807
18 0.184891849147904
19 0.224111870369022
20 0.263525153114856
21 0.303101341806952
22 0.342815771865519
23 0.382647440669031
24 0.422578272191228
25 0.462592676076895
26 0.50267722874539
27 0.542820421124169
28 0.583012445711356
29 0.623245005936686
30 0.663511137461591
31 0.703805036950728
32 0.744121898547246
33 0.784457760947951
34 0.824809368492667
35 0.865174048715874
36 0.905549607268608
37 0.945934239698148
38 0.986326458597033
39 1.02672503413756
40 1.06712894588651
41 1.10753734390832
42 1.14794951739639
43 1.18836486934306
44 1.22878289602259
45 1.26920317029674
46 1.30962532795074
47 1.35004905643008
48 1.39047408547727
49 1.43090017927017
50 1.47132712974206
51 1.5117547508249
52 1.55218287340308
53 1.59261134080024
54 1.63304000464646
55 1.67346872099123
56 1.71389734653856
57 1.75432573488581
58 1.79475373264792
59 1.83518117534279
60 1.87560788290242
61 1.91603365465591
62 1.95645826360456
63 1.99688144977319
64 2.03730291237382
65 2.07772230045403
66 2.1181392016188
67 2.1585531283061
68 2.19896350095657
69 2.23936962723979
70 2.27977067627754
71 2.32016564653491
72 2.36055332573725
73 2.40093224083371
74 2.44130059571187
75 2.48165619415795
76 2.52199634558971
77 2.56231775154316
78 2.60261637193796
79 2.6428872718071
80 2.6831244511457
81 2.72332066201851
82 2.76346721691327
83 2.80355378962229
84 2.84356820479405
85 2.88349620611883
86 2.92332118763705
87 2.96302386840675
88 3.0025818878961
89 3.04196933865093
90 3.08115830055775
91 3.08115830072035
92 3.08115830086114
93 3.08115830097202
94 3.08115830104969
95 3.08115830109812
96 3.08115830112566
97 3.08115830114043
98 3.08115830114809
99 3.081158301152
100 3.08115830115398
101 3.08115830115499
102 3.08115830115549
};
\addlegendentry{$m_\mathrm{B}$}
\end{axis}

\begin{axis}[
axis y line=right,
height=3.5cm,
legend cell align={left},
legend style={
  fill opacity=0.8,
  draw opacity=1,
  text opacity=1,
  at={(0.03,0.03)},
  anchor=south west,
  draw=lightgray204
},
minor xtick={},
minor ytick={},
tick align=outside,
width=7cm,
x grid style={darkgray176},
xlabel={Time since planting in days},
xmajorgrids,
xmin=-5.1, xmax=107.1,
xtick pos=left,
xtick style={color=black},
xtick={-20,0,20,40,60,80,100,120},
y grid style={darkgray176},
ylabel={ \(\displaystyle {\mathrm{\tau}}, {\mathrm{I_{50B}}}\) in \(\displaystyle {\mathrm{^o}}\)C \(\displaystyle {\mathrm{\cdot}}\)d},
ymajorgrids,
ymin=-125, ymax=2125,
ytick pos=right,
ytick style={color=black},
ytick={0,500,1000,1500,2000},
yticklabel style={anchor=west}
]
\addplot [semithick, magenta]
table {%
0 0
1 34.8175291587227
2 69.6350583174455
3 104.452587476169
4 139.270116634892
5 174.087645793616
6 208.905174952341
7 243.722704111066
8 278.540233269793
9 313.357762428521
10 348.175291587251
11 382.992820745983
12 417.810349904717
13 452.627879063448
14 486.306378584327
15 519.935211348939
16 553.487551709475
17 586.879145308816
18 611.979045303264
19 627.968288019068
20 643.660768331774
21 659.230287394575
22 674.730604881779
23 690.186408163304
24 705.611376757343
25 721.013983612544
26 736.399874032668
27 751.773002891135
28 767.136240154076
29 782.491721083187
30 797.841062936293
31 813.185506854743
32 828.526015133785
33 843.863340284453
34 859.198075302593
35 874.530690876813
36 889.861563239369
37 905.190995180737
38 920.519232009623
39 935.846473748289
40 951.172884509647
41 966.498599755583
42 981.823731955622
43 997.148375032359
44 1012.47260788239
45 1027.79649718926
46 1043.12009969166
47 1058.44346403091
48 1073.76663227231
49 1089.0896411738
50 1104.41252325891
51 1119.73530773967
52 1135.058021326
53 1150.38068895217
54 1165.70333444612
55 1181.02598116483
56 1196.34865261695
57 1211.67137309333
58 1226.99416832672
59 1242.31706620328
60 1257.6400975516
61 1272.96329703893
62 1288.2867042108
63 1303.61036471838
64 1318.93433178985
65 1334.25866801769
66 1349.58344755535
67 1364.9087588458
68 1380.23470804425
69 1395.56142335107
70 1410.88906054443
71 1426.2178101024
72 1441.5479064402
73 1456.87963997463
74 1472.21337298119
75 1487.54956056162
76 1502.88877853744
77 1518.23176082494
78 1533.57945001833
79 1548.9330669081
80 1564.29420831162
81 1579.66498957865
82 1595.04826201579
83 1610.44796429845
84 1625.86973097326
85 1641.32203833012
86 1656.81861075619
87 1672.38432240623
88 1688.07375124858
89 1704.06670732518
90 1732.61977275461
91 1761.48723501361
92 1790.35469861619
93 1819.22216415351
94 1848.08963200872
95 1876.95710211164
96 1905.82457398814
97 1934.69204704575
98 1963.55952080161
99 1992.42699494096
100 2021.29446928238
101 2050.16194372792
102 2079.02941822649
};
\addlegendentry{${\mathrm{\tau}}$}
\addplot [semithick, black]
table {%
0 50
1 52.2702053791159
2 54.5404107582322
3 56.810616137349
4 59.0808215164666
5 61.3510268955853
6 63.6212322747054
7 65.8914376538275
8 68.1616430329522
9 70.4318484120803
10 72.7020537912125
11 74.9722591703492
12 77.2424645494887
13 79.5126699286217
14 80.0076184406489
15 80.5025658569502
16 80.9975123021097
17 81.4924577892845
18 81.9874017294263
19 82.4823456271755
20 82.9772895232579
21 83.4722334182195
22 83.9671773123096
23 84.462121205694
24 84.9570650984982
25 85.4520089908227
26 85.9469528827502
27 86.4418967743491
28 86.9368406656765
29 87.4317845567801
30 87.9267284476992
31 88.4216723384663
32 88.9166162291082
33 89.4115601196469
34 89.9065040101005
35 90.4014479004836
36 90.8963917908085
37 91.3913356810851
38 91.8862795713217
39 92.3812234615253
40 92.8761673517016
41 93.3711112418557
42 93.8660551319916
43 94.3609990221132
44 94.8559429122237
45 95.350886802326
46 95.8458306924228
47 96.3407745825165
48 96.8357184726098
49 97.330662362705
50 97.8256062528045
51 98.3205501429111
52 98.8154940330275
53 99.3104379231567
54 99.8053818133022
55 100.300325703468
56 100.795269593658
57 101.290213483878
58 101.785157374132
59 102.28010126443
60 102.775045154777
61 103.269989045183
62 103.76493293566
63 104.25987682622
64 104.754820716879
65 105.249764607655
66 105.74470849857
67 106.23965238965
68 106.734596280927
69 107.229540172439
70 107.72448406423
71 108.219427956358
72 108.714371848891
73 109.209315741912
74 109.704259635524
75 110.199203529855
76 110.694147425062
77 111.189091321343
78 111.684035218942
79 112.178979118165
80 112.673923019393
81 113.1688669231
82 113.663810829876
83 114.158754740457
84 114.653698655764
85 115.148642576959
86 115.643586505528
87 116.138530443405
88 116.633474393191
89 117.128418358645
90 117.623362436476
91 157.416855864166
92 197.210349287806
93 237.003842705615
94 276.797336116512
95 316.590829520838
96 356.384322920094
97 396.177816316036
98 435.971309710042
99 475.764803102993
100 515.55829649539
101 555.351789887503
102 595.145283279471
};
\addlegendentry{${\mathrm{I_{50B}}}$}
\end{axis}

\end{tikzpicture}
	\end{lrbox}
	\begin{figure}[ht]
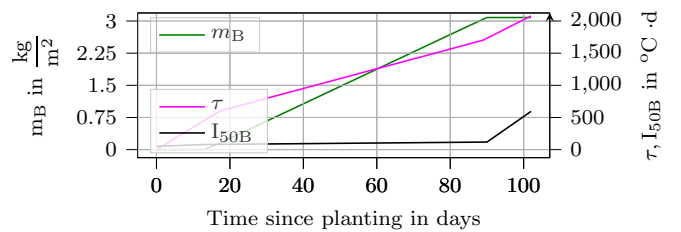

		\centering
		\usebox{\states}
		\caption{Optimal state trajectories of biomass \gls{bmass} (left scale), cumulative temperature \gls{Tcum} and canopy senescence \gls{I50B} (both right scale) for an input optimization with a fixed final time of $N = 102$.}
		\label{fig:states}
	\end{figure}
	
	\newsavebox{\fsolar}
	\begin{lrbox}{\fsolar} 
\begin{tikzpicture}

\definecolor{darkgray176}{RGB}{176,176,176}
\definecolor{orange}{RGB}{255,165,0}

\begin{axis}[
height=3cm,
minor xtick={},
minor ytick={},
tick align=outside,
tick pos=left,
width=8cm,
x grid style={darkgray176},
xlabel={Time since planting in days},
xmajorgrids,
xmin=0, xmax=103,
xtick style={color=black},
xtick={0,20,40,60,80,100,120},
y grid style={darkgray176},
ylabel={\(\displaystyle f_\mathrm{Solar}\ in \ \% \)},
ymajorgrids,
ymin=-4.20209738384976, ymax=99.1858456309764,
ytick style={color=black},
ytick={-20,0,20,40,60,80,100}
]
\addplot [semithick, orange]
table {%
0 5.4430051919775
1 7.53113718384258
2 10.3275966183015
3 13.9959313595053
4 18.6794671689915
5 24.4567694181066
6 31.2880638981949
7 38.9729431769669
8 47.1480871062359
9 55.3435090456707
10 63.0844994969241
11 69.9956895481344
12 75.8621298021751
13 80.6315034892754
14 84.2665034885858
15 87.0624033799522
16 89.1672183703152
17 90.7223130992591
18 91.6108793482964
19 92.0744366395418
20 92.4632263264456
21 92.7923263197593
22 93.0710023418372
23 93.3064704114535
24 93.5047729951435
25 93.6711332961959
26 93.8101419272362
27 93.9258644741894
28 94.0219021658027
29 94.1014247500131
30 94.167191052526
31 94.2215690903073
32 94.2665624261984
33 94.3038441979996
34 94.334796660842
35 94.3605526829052
36 94.3820358939962
37 94.3999972212983
38 94.4150466710653
39 94.4276800900446
40 94.4383011889623
41 94.4472393851448
42 94.4547641104889
43 94.4610962129797
44 94.4664170102858
45 94.4708754665968
46 94.4745938771405
47 94.4776723672099
48 94.4801924468618
49 94.48221980868
50 94.483806512756
51 94.4849926685607
52 94.4858076958561
53 94.4862712246242
54 94.486393675757
55 94.4861765487831
56 94.4856124291583
57 94.4846847147138
58 94.4833670478501
59 94.4816224261142
60 94.4794019479329
61 94.4766431313896
62 94.4732677207119
63 94.469178865983
64 94.4642575245763
65 94.458357885707
66 94.4513015598411
67 94.4428702002259
68 94.4327961331987
69 94.4207504684741
70 94.4063280471761
71 94.389028481308
72 94.3682324788071
73 94.3431726951488
74 94.3128986003106
75 94.2762354181234
76 94.2317381828375
77 94.1776433344069
78 94.111821699659
79 94.0317373727164
80 93.9344157429576
81 93.8164198408986
82 93.673827745715
83 93.5021972429255
84 93.2964996056838
85 93.0510010784532
86 92.759058024521
87 92.4127330271086
88 92.0018546645175
89 91.5099370238688
90 90.4293977785946
91 86.4000876075573
92 79.3464694645451
93 68.2617263708494
94 53.4271821241039
95 37.3134411458303
96 23.3307924024992
97 13.3725080198463
98 7.23498682917428
99 3.7831878870574
100 1.94117280780725
101 0.98569060293458
102 0.497354571369607
};
\addplot [semithick, black, opacity=0.5, dashed]
table {%
0 0.5
1 0.5
2 0.5
3 0.5
4 0.5
5 0.5
6 0.5
7 0.5
8 0.5
9 0.5
10 0.5
11 0.5
12 0.5
13 0.5
14 0.5
15 0.5
16 0.5
17 0.5
18 0.5
19 0.5
20 0.5
21 0.5
22 0.5
23 0.5
24 0.5
25 0.5
26 0.5
27 0.5
28 0.5
29 0.5
30 0.5
31 0.5
32 0.5
33 0.5
34 0.5
35 0.5
36 0.5
37 0.5
38 0.5
39 0.5
40 0.5
41 0.5
42 0.5
43 0.5
44 0.5
45 0.5
46 0.5
47 0.5
48 0.5
49 0.5
50 0.5
51 0.5
52 0.5
53 0.5
54 0.5
55 0.5
56 0.5
57 0.5
58 0.5
59 0.5
60 0.5
61 0.5
62 0.5
63 0.5
64 0.5
65 0.5
66 0.5
67 0.5
68 0.5
69 0.5
70 0.5
71 0.5
72 0.5
73 0.5
74 0.5
75 0.5
76 0.5
77 0.5
78 0.5
79 0.5
80 0.5
81 0.5
82 0.5
83 0.5
84 0.5
85 0.5
86 0.5
87 0.5
88 0.5
89 0.5
90 0.5
91 0.5
92 0.5
93 0.5
94 0.5
95 0.5
96 0.5
97 0.5
98 0.5
99 0.5
100 0.5
101 0.5
102 0.5
};
\end{axis}

\end{tikzpicture}
	\end{lrbox}
	\begin{figure}[ht]
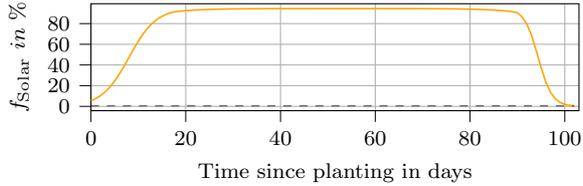

		\centering
		\usebox{\fsolar}
		\caption{Optimal development of $f_{\text{solar}}$ in a growing cycle (yellow). As the biomass directly co-relates with $f_{\text{solar}}$, it is kept to its maximum value for the growth period. Towards the end, it decreases to fulfill the maturity condition $f_{\text{solar}} \leq 0.005$ (dashed line) before harvest.}
		\label{fig:fsolar}
	\end{figure}
	
	\subsectionspace{Optimal control with free final time}
	\vspace{-0.5mm}
	We now extend the optimal control approach to one with free final time, and thus consider the costs and yield over a whole year (as motivated in Fig. \ref{fig:Cropinayear}). The proposed iterative time optimization algorithm in \eqref{eq:costT} and Algorithm \ref{alg} converge to $N = 102$ days after nine iterations, assuming $\delta = 0.01$. The free final time OCP \eqref{eq:costT} converges to the one with a fixed final time \eqref{eq:cost} for $T \approx 1$. The trajectories for the optimal inputs and states are therefore identical to the ones shown in Sec. \ref{sec:fixedfinaltime}. They can be found in Fig. \ref{fig:inputs} and Fig. \ref{fig:states}, respectively. 
	\begin{remark}
	One can see, that for $T \approx 1$ (which is the break condition of Algorithm ~\ref{alg}), \eqref{eq:costT} becomes \eqref{eq:cost} up to a constant factor $\frac{365}{N}$ which does not affect the optimization.
	\end{remark}
	
	\vspace{-0.5mm}
	\subsectionspace{Comparison}
	\vspace{-0.5mm}
    We now assess the efficacy of our approach. The effect on the obtained yield by changing the OCP from a fixed final time \eqref{eq:opt} to a free final time \eqref{eq:optT} is shown in Fig. \ref{fig:yields}. While the yield per growing cycle increases the longer the crop grows before reaching a plateau, the yield per year has a clear maximum yield for a specific length of the growing cycle such that the crop can be harvested multiple times a year, resulting in an overall better performance. 
    
	\newsavebox{\yields}
	\begin{lrbox}{\yields}  
\begin{tikzpicture}

\definecolor{darkgray176}{RGB}{176,176,176}
\definecolor{lightgray204}{RGB}{204,204,204}

\begin{axis}[
height=3.5cm,
legend cell align={left},
legend style={
  fill opacity=0.8,
  draw opacity=1,
  text opacity=1,
  at={(0.09,0.5)},
  anchor=west,
  draw=lightgray204
},
minor xtick={},
minor ytick={},
tick align=outside,
tick pos=left,
width=8cm,
x grid style={darkgray176},
xlabel={Length of growing cycle [d]},
xmajorgrids,
xmin=43.75, xmax=181.25,
xtick style={color=black},
xtick={40,60,80,100,120,140,160,180,200},
y grid style={darkgray176},
ylabel={Yield [\euro/\(\displaystyle m^2\)]},
ymajorgrids,
ymin=3.16676077454022, ymax=72.0526137243964,
ytick style={color=black},
ytick={0,10,20,30,40,50,60,70,80}
]
\addplot [semithick, black]
table {%
50 6.29793590862459
55 7.59384989617708
60 8.88306990137454
65 10.1652599295357
70 11.4393375510103
75 12.7046153800531
80 13.9603967239115
85 15.2064068029657
90 16.4421514389249
95 17.6628053391952
100 18.598003828579
105 19.8267152109117
110 20.6530593284842
115 21.1612245853586
120 21.2833024499915
125 21.3601725266924
130 21.389162620348
135 21.472854997213
140 21.5333701129317
145 21.5312522456825
150 21.5322958584648
155 21.3869089899326
160 21.5222696019067
165 21.47475802686
170 21.4905554308009
175 21.4664516005168
};
\addlegendentry{per cycle}
\addplot [semithick, black, dashed]
table {%
50 45.9749321329595
55 50.3955493109933
60 54.0386752333618
65 57.0818442197005
70 59.6479743731254
75 61.8291281829249
80 63.6943100528464
85 65.2980998009702
90 66.6820586134176
95 67.8623573558553
100 67.8827139743132
105 68.921438590312
110 68.5306059536066
115 67.1638867274424
120 64.7367116187241
125 62.3717037779419
130 60.054187357131
135 58.0562375850575
140 56.1405720801435
145 54.1993591012008
150 52.3952532555977
155 50.3627211698413
160 49.0976775293497
165 47.5047677563872
170 46.141486660249
175 44.7728847667922
};
\addlegendentry{per year}
\end{axis}

\end{tikzpicture}
	\end{lrbox}
	\begin{figure}[ht]
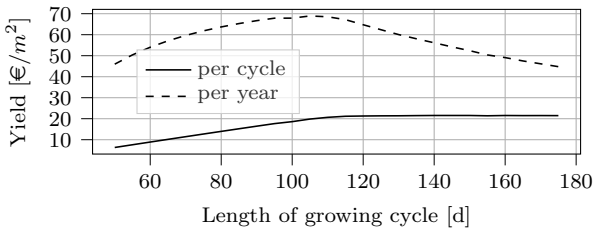

		\centering
		\usebox{\yields}
		\caption{Comparison between yield per year and yield per growing cycle.}
		\label{fig:yields}
	\end{figure}
	
    We also compare our free final time optimal control approach to a scenario where the daily inputs are constant over time. We set the daily input vector \eqref{eq:inputs} to \mbox{$\gls{u} = \begin{bmatrix} 23 \glsunit{T} & 0 \% & 35 \glsunit{SRAD} \end{bmatrix}^\mathsf{T}$}, which are determined to be ideal conditions for wheat when inputs are held constant. Under these conditions, the plant is mature after $N = 110$ days. The biomass obtained after one cycle is $\ensuremath{m_{B}} = 3.08\ \glsunit{bmass}$ and after one year $\ensuremath{m_{B}} = 10.22\ \glsunit{bmass}$. The free final time approach, which uses an optimum of $N = 102$ days and the daily optimal inputs in Fig. \ref{fig:inputs} yields the same biomass after one cycle ($\ensuremath{m_{B}} = 3.08\ \glsunit{bmass}$), but a higher value of $\ensuremath{m_{B}} = 11.02\ \glsunit{bmass}$ for one year. Although they reach the same biomass within a cycle, the input costs for one cycle differ: with 104.55~$\frac{\geneuro}{kg}$ for the optimized daily inputs compared to 149.21 $\frac{\geneuro}{kg}$ for constant ones, the optimized cost per cycle is 30~\% lower. This corresponds to a 25~\% decrease in annual cost while increasing biomass by 8~\%. 
    Finally, we consider a scenario where energy costs are assumed to be zero, to determine the maximum possible biomass yield under the set constraints. We obtain an optimal duration of $N = 119$ and a maximum biomass of $\ensuremath{m_{B}} = 4.09\ \glsunit{bmass}$ per cycle, which is more than 30~\% higher than the current optimum, thus showing the impact of the energy cost to wheat price ratio and also the potential of optimization with respect to profit. 
	
	\sectionspace{Discussion}
	\vspace{-1mm}
	A strength of the presented work is that a model was chosen that can be easily adapted for different crops. Unlike other GH optimizations that use precisely fitted models for tomato or lettuce, with the adapted models (SIMPLE-S \eqref{eq:sys}, -SC \eqref{eq:sysS}, -SCS \eqref{eq:sysT}), it is easy to adjust the parameters and perform optimization for 12 more crops, even though in this work, we only show the optimization for one crop as an example. However, since we did not perform real experiments with the new models, but only compared them with the results of the original SIMPLE model \citep{Zhao.2019}, the models are at most as good as the SIMPLE model itself. Also, since the SIMPLE model, like most other crop models, was developed based on data from field cultivation, the growth conditions of the new models are not yet adapted to VFs. This limits the choice of inputs by avoiding areas for which the model was not designed, i.e., areas that do not occur in nature. Adapting the models to VFs, thus allowing higher input values, may lead to even better optimization results.
    
    \vspace{-0.5mm}
    In the optimization section, we show that both proposed approaches for the OCPs \eqref{eq:opt} and \eqref{eq:optT} perform well as they solve the problems even with more input parameters than in other studies. The optimal inputs remain within the given constraints and show a smooth trajectory that leads to a mature plant after the given time. It is also shown that the annual gain depends on the growth time of the plant. This confirms that for optimal plant growth, the growing time should not be determined manually, but must be part of the optimization in order to make a meaningful statement about the highest possible profit per year.
    However, a major problem with choosing wheat as an example, which we have already pointed out and which is discussed in detail by \citet{asseng2020}, is that growing wheat in a VF is not (yet) profitable. In our paper, we assume a wheat price 400 times higher than current market prices, since the optimization otherwise yields a result of zero. Our price assumptions are based on current average global energy prices, but these are about 10 times higher than the best possible global conditions assumed in \citep{asseng2020}. Under these conditions, they conclude that wheat prices would need to be about 50 times higher than current prices to achieve profitability in a VF. Although the energy cost factors in this paper are only an estimate, our results are of a similar order of magnitude when the energy cost difference of a factor of 10 is included. Thus, it is clear that cultivation only makes sense in special cases (e.g., cultivation in space or in very remote areas), that the energy efficiency of VFs needs to be significantly increased (e.g., through local production and use of renewable energy such as solar power and wind, and closed cycles for water and CO\textsubscript{2}), and that VFs for wheat would need to be subsidized, as is common for outdoor cultivation. Feasibility is more likely for other crops (tomato, lettuce, cotton, etc.) that can be optimized with the same approach and also with the same model.
	
	\sectionspace{Conclusion}
	In this work, we present an approach to make VFs more energy efficient by determining the daily optimal conditions for the plant. Our results show that the models \mbox{SIMPLE-S}~\eqref{eq:sys} and -SC~\eqref{eq:sysS} introduced for this purpose, which we use in a state-space form to apply control theory to them, reproduce the fitted data of the original SIMPLE model \citep{Zhao.2019}. We also show that our proposed optimal control algorithms with fixed and free final times are feasible for optimizing multiple input variables and improve the efficiency of plant growth in a VF. Considering the growth duration leads us to a new cost function that gives better results in terms of yield and energy cost, but makes the length of the input vector variable. For this reason, the free final time algorithm uses a new model definition, SIMPLE-SCS \eqref{eq:sysT}, in combination with an iterative approach to solve the OCP. In this example for wheat, the optimization shows a 25 \% reduction in energy cost and an increase in harvest per year compared to inputs considered ideal for the crop, but constant for each day. Despite some limitations of the model and the current feasibility gap for wheat in VFs due to high energy costs, the results show the potential of using optimal control in VFs with even more sophisticated approaches than those already used for GHs. To the best of our knowledge, our study is the first to optimize the three crop inputs temperature, artificial radiation, and irrigation in a VF, explicitly accounting for crop growth time and maturity.
	Future work will include analysis and fitting of other models and constraints that can be used in control theory. This will allow for other parameters to be considered in a VF, such as soil nutrients and also other crops that are likely to be used in VFs in the near future. Additionally, we will extend the open-loop OC approach to a closed-loop version to better account for model inaccuracies and other external perturbations.

\bibliography{Literature}             
\end{document}